\documentclass[
amsmath,amssymb,aps,
author-numerical,
twocolumn,
nofootinbib,
superscriptaddress,
]{revtex4-2}

\usepackage{graphicx}
\usepackage{dcolumn}
\usepackage{bm}
\usepackage{xcolor}
\usepackage[normalem]{ulem}
\usepackage{hyperref}

\newcommand{\nhat}{\hat{\mathbf{n}}}
\newcommand{\lvec}{\mathbf{l}}

\newcommand{\Lvec}{\mathbf{L}}

\newcommand{\dtwol}[1]{\frac{d^2 \lvec_{#1}}{(2\pi)^2}}

\def\muK{\mu{\mbox{K}}}

\newcommand{\nmaps}{n_\mathrm{maps}}

\newcommand{\tsurvey}{\Delta t_\mathrm{survey}}
\newcommand{\tcadence}{\Delta t_\mathrm{cadence}}

\mathchardef\mhyphen="2D

\definecolor{darkred}{RGB}{150,50,50}

\newcommand{\SMU}{Department of Physics,
Southern Methodist University, 
Dallas, TX 75275, USA}

\newcommand{\ASU}{School of Earth and Space Exploration,
Arizona State University, 
Tempe, AZ 85287, USA}

\begin{document}


\title{Time-dependent deflection reconstruction: new technique to search for \\
gravitational waves with the cosmic microwave background}

\author{Alvin~Leluc}
\affiliation{\SMU}
\author{Joel~Meyers}
\affiliation{\SMU}
\author{Alexander~van~Engelen}
\affiliation{\ASU}

\date{\today}
 
\begin{abstract}
Gravitational waves (GWs) passing through the Earth cause a correlated pattern of time-dependent deflections of the apparent position of astronomical sources.
We build upon standard lensing reconstruction techniques to develop a new time-dependent optimal quadratic estimator, providing a novel technique to search for the deflections produced by GWs using observations of the cosmic microwave background (CMB).
We find that the time-dependent deflection reconstruction is many orders of magnitude more sensitive than the ordinary static lensing estimator, and that it can be employed with the data collected by existing and future CMB surveys, without requiring any modification to the experimental or survey design.
We demonstrate that CMB surveys offer sensitivity to GWs across a broad frequency range: while the sensitivity  will not be competitive over the frequency range covered by pulsar timing arrays, the  coverage band extend to both lower and higher frequencies.  
Finally, we discuss how our methods can be adapted to search for other time-varying signals, and also how they can be applied to surveys at other wavelengths.
 \end{abstract}

\maketitle

\section{Introduction}

Observations of gravitational waves (GWs) provide a valuable source of information about the Universe, complementary to electromagnetic observations~\cite{Maggiore:2018sht,Bailes:2021tot}. 
Several techniques have emerged in the hunt for GWs across a range of frequencies~\cite{Maggiore:2007ulw}. 
Ground-based laser interferometers like LIGO~\cite{LIGOScientific:2014pky}, Virgo~\cite{VIRGO:2014yos}, and KAGRA~\cite{KAGRA:2020tym} provide sensitivity to GWs between $10^1$ and $10^4$~Hz~\cite{KAGRA:2013rdx,KAGRA:2023pio}. 
Future space-based laser interferometers like LISA~\cite{LISA:2024hlh} will probe the range $10^{-4}$ to $10^0$~Hz \cite{GairEtAl:SpacebasedGravitational2021}. 
At lower frequencies, pulsar timing arrays (PTAs) are used to explore the $10^{-9}$ to $10^{-7}$~Hz regime \cite{NANOGrav:2023hfp,EPTA:2023fyk,Antoniadis:2022pcn,Xu:2023wog}. 
Searches are actively underway for traces of horizon-scale primordial GWs which leave characteristic imprints in the polarization of the cosmic microwave background (CMB)~\cite{KamionkowskiEtAl:StatisticsCosmic1997,ZaldarriagaSeljak:AllSkyAnalysis1997,Kamionkowski:2015yta}.

Another means to search for GWs is provided by precise astrometric measurements. 
GWs passing through Earth cause deflections of the apparent source positions of incoming radiation. 
It has been shown that monitoring the precise positions of distant astronomical objects on the sky thereby enables sensitivity to GWs \cite{BraginskyEtAl:PropagationElectromagnetic1990,PyneEtAl:GravitationalRadiation1996,GwinnEtAl:QuasarProper1997,KaiserJaffe:BendingLight1997,BookFlanagan:AstrometricEffects2011}. 
Absolute astrometric measurements from observatories like \textsl{Gaia}~\cite{Gaia:2016zol} can provide sensitivity to GWs comparable to that achieved by PTAs~\cite{BookFlanagan:AstrometricEffects2011,MooreEtAl:AstrometricSearch2017,Darling:2018hmc,Jaraba:2023djs,Darling:2024myz} within a similar range of frequencies.
Recent investigations have shown how relative astrometric measurements of distant galaxies~\cite{Wang:2020pmf,Wang:2022sxn,Pardo:2023cag} can also be used to search for GWs.

The GW-induced deflection effect remaps observations of points on the sky in a way that is entirely analogous to gravitational lensing by large-scale structure, except the GW effect varies on much shorter timescales~\cite{BookFlanagan:AstrometricEffects2011,MignardKlioner:AnalysisAstrometric2012,MooreEtAl:AstrometricSearch2017,Qin:2018yhy,Pardo:2023cag,Caliskan:2023cqm,MentastiContaldi:CosmicShimmering2024}. 
As a result, all of the tools that are used to search for the effects of gravitational lensing deflection can be adapted to the search for GWs.
For example, GWs induce varying ellipticity of galaxies allowing for time-resolved cosmic shear measurements to be used in the search for GWs~\cite{MentastiContaldi:CosmicShimmering2024}.

The science goals of current and future CMB surveys~\cite{CMB-S4:2016ple,SimonsObservatory:2018koc,Abazajian:2019eic,Sehgal:2019ewc,SimonsObservatory:2025wwn} include the search for new light degrees of freedom, measurement of cosmological structure growth with CMB lensing and galaxy clusters, and the observation of variable and transient astrophysical sources. These scientific targets drive the design toward wide sky coverage (roughly 60\% of the sky), fine angular resolution (arcminute-scale), and high-cadence observations (covering large sky areas each day).
These same design features naturally offer advantages when using CMB surveys to search for the time-dependent imprints of GWs.
Techniques developed to search for gravitational lensing of the CMB~\cite{HuOkamoto:MassReconstruction2002,OkamotoHu:CMBLensing2003,LewisChallinor:WeakGravitational2006} have been optimized over the past several years and applied for higher precision measurements of lensing by large-scale structure~\cite{Planck:2018lbu,ACT:2023kun,SPT-3G:2024atg}.
However, none of these CMB tools have addressed the time-dependent deflection signatures expected from GWs, as employed in astrometric GW searches that utilize point sources~\cite{Darling:2018hmc,Jaraba:2023djs}.

\begin{figure*}
    \centering
    \includegraphics[width=\textwidth]{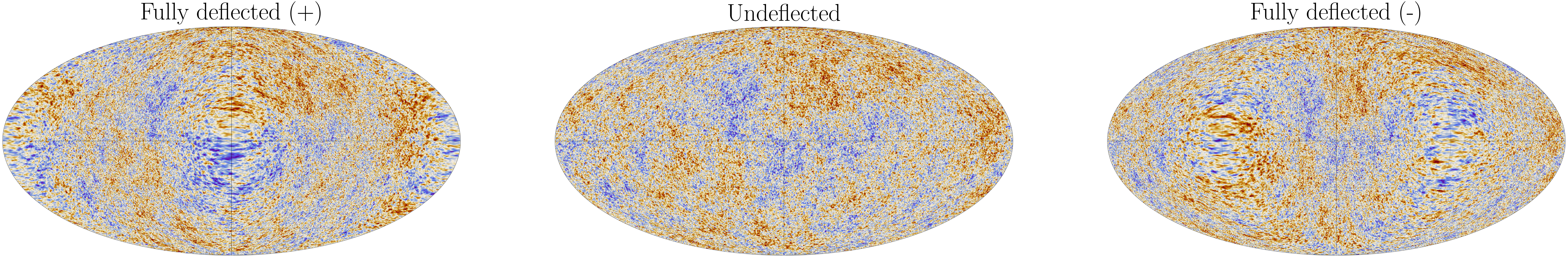}
    \caption{
    Illustration of the (exaggerated) effect on the observed CMB due to gravitational waves passing through Earth, modeled here with divergence-type quadrupolar deflection potential $\phi_{l=2,m=\pm2}$.  The stretching and squeezing  of the CMB hot and cold spots is apparent.
    While a continuous gravitational wave at a single temporal frequency would cause the deflections to oscillate in time with a fixed spatial distribution in this way, a stochastic background of gravitational waves would produce a random time-dependent pattern of deflections with statistics governed by Eq.~\eqref{eq:thetaQrmsl}.
    }
    \label{fig:3PanelCMB}
\end{figure*}

In this paper, we will develop a new estimator designed to search for time-dependent deflection of CMB anisotropies, such as those imprinted by GWs. 
Most cosmological analyses of CMB data are performed with CMB maps constructed from averages taken over the full survey duration. However, the signal of interest for GW searches varies over time, thereby motivating the analysis of a sequence of time-ordered CMB maps.
We will show how applying a Fourier transform in the temporal domain  to a time series of CMB maps allows us to construct an estimator for time-dependent deflections across a range of frequencies with convenient noise properties that can be used to isolate the subtle, time-varying deflections induced by GWs.
We will then forecast the sensitivity of our new estimator to a stochastic GW background (SGWB) when using current and next-generation CMB surveys.
We briefly comment on how this new framework can be extended to different cosmic surveys and to different targets.

\section{Impact of Gravitational Waves on the CMB}
\label{sec:GW_CMB}

We start by considering the deflections produced by GWs and how they differ from gravitational lensing by large-scale structure in a few key aspects. 
First, the pattern of deflections due to GWs is expected to produce equal power in divergence-type ($E$-mode) and curl-type ($B$-mode) deflection~\cite{BookFlanagan:AstrometricEffects2011}, whereas gravitational lensing due to large-scale structure only produces divergence-type lensing (except on small angular scales where post-Born lensing corrections can become relevant~\cite{Pratten:2016dsm,Marozzi:2016uob,RobertsonEtAl:DetectableSignals2025}). 
Next, the deflection due to GWs peaks on large angular scales and falls rapidly with multipole, reflecting the primarily quadrupolar nature of the deflection. 
This characteristic angular dependence of GW deflection is analogous to the Hellings-Downs curve, which describes the spatial correlation of pulsar timing residuals in the presence of GWs ~\cite{HellingsDowns:UpperLimits1983}.
Finally, and most essentially, the deflection due to a SGWB is time dependent, while the lensing by large-scale structure remains static over the duration of the survey, changing only on cosmological time scales (and similarly the primary CMB anisotropies evolve very slowly compared to the survey duration~\cite{Lange:2007tx,Zibin:2007mu,Moss:2007bu}).  
Figure~\ref{fig:3PanelCMB} shows the effect of GWs on a simulated map of the CMB.

It was shown in Ref.~\cite{BookFlanagan:AstrometricEffects2011} that a statistically isotropic SGWB produces random angular deflections of the apparent positions of distant astronomical sources with variance, in radians, of 
\begin{equation}
    \bar\theta_\mathrm{rms}^2 = \frac{1}{4\pi^2} \int d\log f \left(\frac{H_0}{f}\right)^2 \Omega_\mathrm{GW}(f) \, ,
    \label{eq:thetabarrms}
\end{equation}
where $H_0$ is the Hubble constant and $\Omega_{\mathrm{GW}}(f)$ is the energy density in GWs per unit logarithmic frequency divided by the critical energy density, i.e.~$\bar\Omega_{\mathrm{GW}}=\int d \log f \,  \Omega_{\mathrm{GW}}(f)$, and we use an overbar to denote quantities integrated over frequency (such that $\bar\theta_\mathrm{rms}^2 = \int d\log f \, \theta_\mathrm{rms}^2(f)$).
The derivation of Eq.~\eqref{eq:thetabarrms} presented in Ref.~\cite{BookFlanagan:AstrometricEffects2011} proceeds by first computing the apparent deflection of a photon source due to an arbitrary GW signal $h_{ij}$, assuming that both the source and observer a freely falling in an expanding universe.  Under the assumption that the photon source is at a distance much larger than the wavelength of the GW and that the GW wavelength is short compared to the horizon size, the deflection takes a simple form for plane GWs, depending only on the GW evaluated at the position of the observer
\begin{equation}
    \delta n^i(\tau,\nhat) = \frac{n^i+p^i}{2(1+\mathbf{p}\cdot\nhat)}h_{jk}(\tau,0)n_jn_k-\frac{1}{2}h_{ij}(\tau,0)n_j \, ,
    \label{eq:GWdeflection}
\end{equation}
where $\nhat$ denotes the direction in the sky, $\tau$ is the time of observation, the GW propagates along direction $\mathbf{p}$, and $h_{ij}(\tau,0)$ refers to the GW at the position of the observer.  The apparent deflection of the source accounts for both the perturbed path of the photon and the perturbation to the observer's local proper reference frame in the presence of the GW.  The statistical properties of the deflection due to a SGWB can then be computed from the two-point correlation function of the metric perturbations, which can be treated as a superposition of plane waves.

The deflections produced by these GWs can be broken down into harmonic modes  $L$, divergence- and curl-type deflections $\Gamma\in[\phi,\psi]$, and GW temporal frequencies $f$ according to 
\begin{align}
    \theta_{\mathrm{rms}}^2(f) 
    &= \sum_{L=2}^\infty \sum_{\Gamma=\phi,\psi} \theta_{L,\Gamma}^2(f) \nonumber \\
    &=\sum_{L=2}^\infty \sum_{\Gamma=\phi,\psi}\frac{g_{\Gamma}\, \alpha_L^{\Gamma \Gamma}\,H_0^2\,\Omega_{\mathrm{GW}}(f)}{4\pi^2\,f^2} \, ,
    \label{eq:thetaQrmsl}
\end{align}
with $g_\phi=g_\psi=1/2$ and $\sum_L \alpha_L^{\Gamma\Gamma}=1$.
The deflection power is restricted to the largest angular scales, with a quadrupole contribution of $\alpha_2\simeq0.83$, an octopole contribution of $\alpha_3\simeq0.12$, and trace contributions at higher $L$~\cite{BookFlanagan:AstrometricEffects2011}. 
While the focus of Ref.~\cite{BookFlanagan:AstrometricEffects2011} was on surveys providing precise astrometric measurements of point-like sources, the deflection power expressed in Eq.~\eqref{eq:thetaQrmsl} is a general deflection that affects our entire observed sky, indicating that any astronomical survey could in principle be used to search for these deflections, including CMB surveys (see also Refs.~\cite{Madison:2020atu,Madison:2020xhh,Zwick:2024hfn} for a similar construction applied to GW memory effects).

The time-dependent nature of these GW-induced deflections motivates an analysis of CMB surveys that takes account of the time series of observations, rather than relying only on a single map constructed from the entire survey duration.  
Commonly, given a survey duration of $\tsurvey$, typically several years, one constructs a single observed CMB map that represents the average over the entire survey,  $T^\mathrm{obs}(\nhat)$, with $\nhat$ the direction of observation. 
We can also use the time-ordered data to form 
$\nmaps=\tsurvey/\tcadence$
separate maps $T(\nhat,t_j)$. 
Each map in this set will be based on observations taken during a limited time interval $\tcadence$, labeled $\{t_j\}_{j=0}^{\nmaps - 1}$. 
For instance, $T(\nhat,t_j)$ could represent a map for each day or each week of  data. 
This approach is commonly taken in power spectrum analysis, in order to take cross-spectra between such maps without incurring any noise bias~\cite{Planck:2019nip,ACT:2025fju,SPT-3G:2025bzu}, as well as for so-called ``cross-split" estimators in CMB lensing reconstruction \cite{Madhavacheril:2020ido}.  

Each of these maps can be written as the sum of the components that do not vary over the duration of the survey and those that do, i.e. in the absence of GWs, we have $ T^\mathrm{obs}(\nhat,t_j) = T^\mathrm{CMB}(\nhat) + T^\mathrm{N}(\nhat,t_j)$.
Here, $T^\mathrm{N}$ denotes the noise realization for the given map. Astrophysical foregrounds would, like the CMB, appear as a time-independent contribution to the observed temperature, and the noise level of the individual maps will be much larger than that of the full survey.

\begin{figure}
    \centering
    \includegraphics[width=\columnwidth]{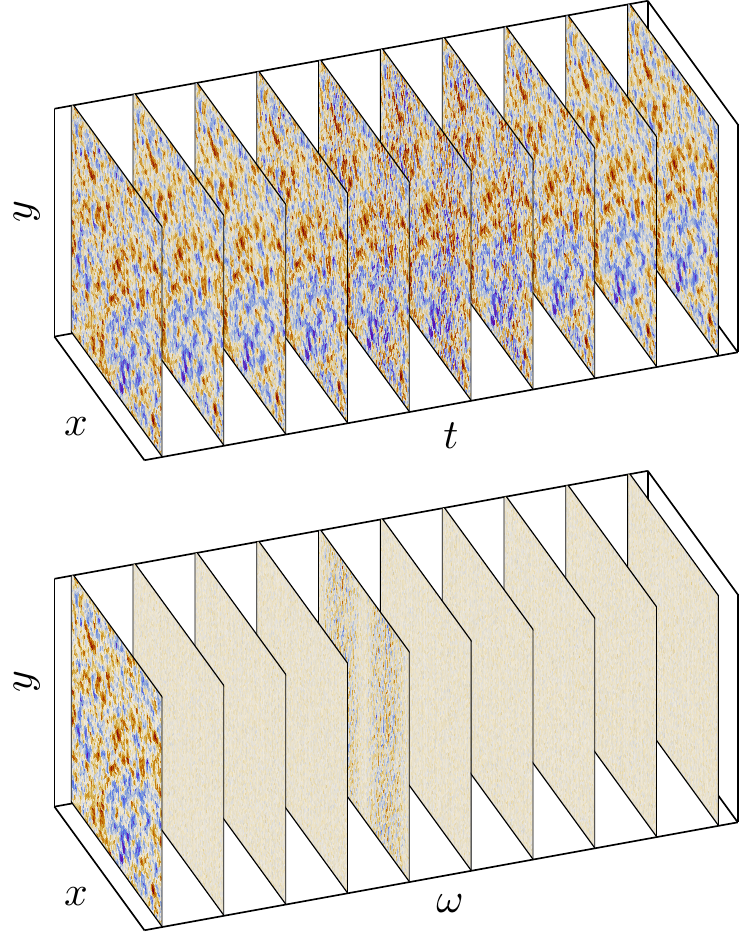}
    \caption{\textit{Top}: A time series of CMB maps constructed from a set of fixed-interval observations. \textit{Bottom}: A Fourier transform of the time series above, which isolates the primary, constant CMB to the $\omega=0$ slice.  The higher frequency modes normally consist only of noise, at the same depth.  A time-dependent GW deflection of the CMB maps  has been injected in the fourth nonzero frequency slice (significantly exaggerated for visibility), showing the 'echo' generated at that slice (Eq.~\eqref{eq:deflectedT}).  Our estimator searches for correlations between this echo and the constant slice.}
    \label{fig:Cartoon}
\end{figure}

Now we apply a Fourier transform in both the spatial and temporal domains, working with continuous transforms for simplicity
\begin{equation}
    T^\mathrm{obs}(\nhat, t) = \int \frac{d\omega}{2\pi} \int \frac{d^2\lvec}{(2\pi)^2} \mathcal{T}^\mathrm{obs}(\lvec,\omega) e^{i\lvec \cdot \nhat}e^{i \omega t} \, ,
    \label{eq:Fourier Transform}
\end{equation}
and we use script symbol $\mathcal{T}(\lvec,\omega)$ to highlight that we have taken the three-dimensional Fourier transform over the two spatial dimensions as well as in the time dimension.  This is to be distinguished from the usual two-dimensional Fourier transform of the CMB temperature, defined by $T(\nhat) = \int \frac{d^2\lvec}{(2\pi)^2} T(\lvec) e^{i\lvec \cdot \nhat}$.
The advantage of the temporal Fourier transform is that our $\omega = 0$ slice will contain contributions from the CMB and foreground fluctuations 
$\mathcal{T}^\mathrm{obs}(\lvec,\omega=0)=\tsurvey (T^\mathrm{CMB}(\lvec)+T^\mathrm{N}(\lvec))$,
but at $\omega \ne 0$, only signals that vary at the given temporal frequency $f = \omega /2\pi$ and instrumental noise will contribute.
This is visually represented in Figure~\ref{fig:Cartoon}.
If there is no appreciable time variation in any of the CMB, foregrounds, or the instrumental systematics, then the $\omega \neq 0$ components will only contain time-varying signals and noise with no contribution from the primary CMB or astrophysical foregrounds.

We assume that the noise is stationary in both space and time such that the noise power is diagonal in both the spatial and temporal Fourier domains
\begin{align}
  &\langle \mathcal{T}^\mathrm{N}(\lvec_1, \omega_1) \mathcal{T}^\mathrm{N}(\lvec_2, \omega_2)^\star \rangle  \nonumber \\
  & \qquad = (2 \pi)^3 \mathcal{N}_{l_1,\omega_1}^{\mathcal{TT}} \delta(\omega_1 - \omega_2) \delta^{(2)}(\lvec_1 - \lvec_2) \, ,
  \label{eq:NoisePower}
\end{align} 
where again the script $\mathcal{N}_{l_1,\omega_1}^\mathcal{TT}$ refers to the 3d noise power spectrum.
We do not assume the noise power 
$\mathcal{N}_{l,\omega}^\mathcal{TT}$
to be constant as a function of $l$, meaning that there may be excess fluctuations at large scales due to, e.g., atmospheric noise fluctuations. However, we do assume, for simplicity, that the noise is white and stationary in the temporal domain, so that each span of observations will have the same noise statistics as each other span taken at different times. 
Therefore, all temporal Fourier modes will contain the same noise power as at $\omega = 0$, namely
$\mathcal{N}_{l,\omega}^\mathcal{TT}$,
but  the time-independent CMB and foregrounds will appear only at the $\omega = 0$ mode.
Note that the 3d noise power spectrum for temperature fluctuations at a particular temporal frequency 
$\mathcal{N}_{l,\omega}^\mathcal{TT}$
is related to the 2d noise power for the full duration of the survey $N_l^{TT}$ by  
$\mathcal{N}_{l,\omega}^\mathcal{TT}=2\pi \delta(0)N_{l}^{TT}$
where we interpret the delta function as $2\pi \delta(0)=\tsurvey$.
In particular, we define the (beam-deconvolved) 2d noise power that we use throughout the paper as
\begin{align}
    N_l^{TT} &= \Delta_T^2\exp\left(l(l+1)\frac{\theta^2_\mathrm{fwhm}}{8\log 2}\right) \, ,
\end{align}
where $\Delta_T$ is the survey depth and $\theta_\mathrm{fwhm}$ is the full-width at half-maximum of the beam.  This is related to the 3d noise power by a factor of $\tsurvey$, which takes the form
\begin{align}
    \mathcal{N}_{l,\omega}^\mathcal{TT} &= \tsurvey \,\Delta_T^2  \exp\left(l(l+1)\frac{\theta_\mathrm{fwhm}^2}{8\log 2}\right) \nonumber \\
    &= (\mathrm{NET}_\mathrm{survey})^2 (4\pi f_\mathrm{sky}) \exp\left(l(l+1)\frac{\theta_\mathrm{fwhm}^2}{8\log 2}\right) \, ,
\end{align}
where $\mathrm{NET}_\mathrm{survey}$ is the noise-equivalent temperature of the survey in the time domain, also known as the mapping speed or the instantaneous sensitivity of the detector array; and $f_\mathrm{sky}$ is the fraction of sky covered by the survey.
For example, if 5 years of CMB-S4 observations result in a total survey depth of 1~$\mu$K-arcmin (i.e.~2d white noise power on large scales  of $N_l^{TT} = (1~\muK\mhyphen  \mathrm{arcmin})^2$) the 3d, or instantaneous, noise power  of this survey would be 
$\mathcal{N}_{l,\omega}^\mathcal{TT}=5~(\muK\mhyphen \mathrm{arcmin})^2\,\mathrm{yr}$, or $\mathcal{N}_{l,\omega}^\mathcal{TT}=1.6\times10^8~(\muK\mhyphen \mathrm{arcmin})^2/\mathrm{Hz}.$
We can therefore write the total observed 3d CMB power spectrum in terms of the 2d temperature and noise spectra, assuming no time-dependent signals, as
\begin{equation}
  \mathcal{C}_{l,\omega}^{\mathcal{TT},\mathrm{obs}} = 2\pi \delta(\omega) C_{l}^{TT,\mathrm{CMB}}  +  2\pi \delta(0) N_l^{TT} \, .
  \label{eq:CTTobs}
\end{equation}

A CMB survey thus effectively contains not only one map, or slice, at survey depth, but $n_\mathrm{maps} = \tsurvey/\tcadence$ such slices across the temporal frequency domain---each with the same noise level.  Given that maps can be constructed on an hourly or daily basis, and that surveys can run for five years or more, this can yield many hundreds or thousands of maps.  However, with CMB studies, one typically only constructs the time-averaged map, which corresponds to the $\omega=0$  slice, and ignores the other slices.  In the absence of time-dependent effects, the CMB and foregrounds are limited to the slice at $\omega=0$.  The slices at $\omega>0$ enable us to search for signatures of statistical changes over the duration of the survey.

\section{Time-Dependent Deflection Reconstruction}
\label{sec:Time-Dependent_Recon}

Let us start by supposing that there is a GW passing through the Earth.  
This will manifest as a time-varying angular deflection vector $\mathbf{d}(t)$ that can be represented in terms of a scalar lensing potential $\phi(t)$ and  lensing cross-potential $\psi(t)$, such that $\mathbf{d}(\nhat, t) = \nabla \phi(\nhat,t) + \star\nabla \psi(\nhat, t)$. 
Here, the $\star$ is a ninety-degree rotation operator such that $\star \hat{x}=\hat{y}$ and $\star\hat{y} = -\hat{x}$~\cite{Hirata:2003ka, CoorayEtAl:CosmicShear2005}.
The deflected CMB temperature anisotropies are then given by
\begin{equation}
    T(\nhat, t) =   \tilde T\left[\nhat + \nabla \phi(\nhat, t) + \star\nabla \psi(\nhat, t) \right] \, ,
\end{equation}
where $\tilde T$ is the undeflected CMB temperature map.
From here we can express our deflected map to first order in $\phi$ and $\psi$  in Fourier space as discussed in the previous section 
\begin{align}
    &\mathcal{T}(\lvec,\omega) = \tilde{\mathcal{T}}(\lvec,\omega)  \nonumber \\
    &-  \int \frac{d\omega'}{2\pi} \int \frac{d^2l'}{(2\pi)^2}  
    \lvec'\cdot(\lvec-\lvec')\,\tilde{\mathcal{T}}(\lvec',\omega') \phi(\lvec-\lvec',\omega-\omega') \nonumber \\
    &-  \int \frac{d\omega'}{2\pi} \int \frac{d^2l'}{(2\pi)^2}   
    \lvec'\cdot \star(\lvec-\lvec')\,\tilde{\mathcal{T}}(\lvec',\omega') \psi(\lvec-\lvec',\omega-\omega')\, ,
    \label{eq:deflectedT}
\end{align}
where in the case that the undeflected temperature does not depend on time, we have $\mathcal{T}(\lvec,\omega) = 2\pi\delta(\omega)T(\lvec)$.

As in the case of standard CMB lensing, we can construct a quadratic estimator for time-dependent deflection by using the correlation that is induced between different Fourier modes. Due to the time dependence of the signal of interest, we consider the correlation at different values of $\mathbf{l}$ and $\omega$. When averaged over realizations of the primary CMB temperature fluctuations and noise, the off-diagonal ($\lvec_1 \neq -\lvec_2$) 
two-point function  is
\begin{align}
    &\langle \mathcal{T}(\lvec_1,\omega_1) \mathcal{T}(\lvec_2,\omega_2) \rangle_\mathrm{CMB} \nonumber \\
    & \qquad =  \sum_\Gamma f^{\Gamma}(\lvec_1,\lvec_2,\omega_1,\omega_2)  \Gamma(\lvec_1 +\lvec_2, \omega_1+\omega_2)  
    \label{eq:Tlen2pt}
\end{align}
where we defined the 3d mode couplings
\begin{align}
    f^\phi(\lvec_1,\lvec_2,\omega_1,\omega_2) &= \lvec_1\cdot(\lvec_1+\lvec_2)\,\tilde{\mathcal{C}}_{l_1,\omega_1}^\mathcal{TT} \nonumber \\
    & \hphantom{{}=} +\lvec_2\cdot(\lvec_1+\lvec_2)\,\tilde{\mathcal{C}}_{l_2,\omega_2}^\mathcal{TT} \\
    f^\psi(\lvec_1,\lvec_2,\omega_1,\omega_2) &= \lvec_1\cdot\star(\lvec_1+\lvec_2)\,\tilde{\mathcal{C}}_{l_1,\omega_1}^\mathcal{TT} \nonumber \\
    & \hphantom{{}=} +\lvec_2\cdot\star(\lvec_1+\lvec_2)\,\tilde{\mathcal{C}}_{l_2,\omega_2}^\mathcal{TT} \, .
\end{align}
Assuming that the undeflected temperature map is time-independent, with angular (2d) power spectrum $\tilde C_l^{TT}$ (so that $\tilde{\mathcal{C}}_{l,\omega}^\mathcal{TT} = 2\pi \delta(\omega)\tilde C_l^{TT}$), we can 
construct our quadratic estimator for $\Gamma$ as
\begin{align}
    &\hat{\Gamma}(\Lvec, \Omega) 
    = A^\Gamma(L,\Omega) \nonumber \\
    &\qquad \times \int \dtwol{1}  \mathcal{T}^\mathrm{obs}(\lvec_1, 0) \mathcal{T}^\mathrm{obs}(\lvec_2, \Omega)
    F^\Gamma(\lvec_1, \lvec_2, \Omega) \, ,
    \label{eq:estimator0Omega}
\end{align}
where we defined $\lvec_2=\Lvec -\lvec_1  $. The estimator defined in Eq.~\eqref{eq:estimator0Omega}, which correlates $\mathcal{T}(\lvec_1,0)$ with $\mathcal{T}(\lvec_2,\Omega)$, amounts to looking for a deflected image of the time-averaged temperature map at the frequency $\Omega$ of a passing gravitational wave.  This is shown graphically in Figure~\ref{fig:Cartoon}. 

Following, e.g., Ref.~\cite{HuOkamoto:MassReconstruction2002}, we require that the estimator is unbiased, $\left\langle \hat\Gamma(\Lvec,\Omega)\right\rangle_\mathrm{CMB} = \Gamma(\Lvec,\Omega)$ which fixes the normalization  
\begin{align}
  A^\Gamma(L,\Omega) = \left[\int \dtwol{1} f^\Gamma(\lvec_1,\lvec_2,0,\Omega) F^\Gamma(\lvec_1, \lvec_2, \Omega)\right]^{-1} \, .
  \label{eq:EstimatorNormalization}
\end{align}
%
The variance of the estimator is
\begin{align}
    \left\langle \hat{\Gamma}(\Lvec,\Omega) \hat{\Gamma}(\Lvec',\Omega')^\star\right\rangle &=(2\pi)^3 \delta^{(2)}(\Lvec-\Lvec')\delta(\Omega-\Omega') \nonumber \\
    & \quad \times \left( \mathcal{C}^{\Gamma\Gamma}_{L,\Omega}+\mathcal{N}^{\Gamma\Gamma}_{L,\Omega} \right) \, ,
\end{align}
and the choice of filter that minimizes the variance  is
\begin{align}
    F^\Gamma(\lvec_1, \lvec_2, \Omega) &
    = \frac{ f^\Gamma(\lvec_1,\lvec_2,0,\Omega)}{ \mathcal{C}_{l_1,0}^{\mathcal{TT},\mathrm{obs}} \mathcal{C}_{l_2,\Omega}^{\mathcal{TT},\mathrm{obs}}} \, .
\end{align}
The 3d noise power of our time-dependent deflection reconstruction is then
\begin{align}
    \mathcal{N}^{\Gamma\Gamma}_{L,\Omega} &= 2\pi\delta(0) A^\Gamma(L,\Omega) \nonumber \\
    &= 2\pi\delta(0) \left[ \int \dtwol{1} \left(\lvec_1 \odot \Lvec \right)^2 \, \frac{\left(\tilde{\mathcal{C}}^\mathcal{TT}_{l_1,0}\right)^2}{ \mathcal{C}_{l_1,0}^{\mathcal{TT},\mathrm{obs}} \mathcal{C}_{\left|\Lvec-\lvec_1\right|,\Omega}^{\mathcal{TT},\mathrm{obs}}}  \right]^{-1} \nonumber \\
    &= \tsurvey\left[ \int \dtwol{1} \left(\lvec_1 \odot \Lvec \right)^2 \, \frac{\left(\tilde C^{TT}_{l_1}\right)^2}{ C_{l_1}^{TT,\mathrm{obs}} N_{\left|\Lvec-\lvec_1\right|}^{TT}}  \right]^{-1} \, ,
    \label{eq:Nphi=NPsi}
\end{align}
where 
we assumed $\Omega\neq0$ (and again used that $2\pi\delta(0)=\tsurvey$). 
We have introduced the operator $\odot$ to represent a dot product when $\Gamma = \phi$ and a dot product with the ninety-degree rotated vector when $\Gamma=\psi$~\cite{CoorayEtAl:CosmicShear2005}.
Note that if we had taken $\Omega=0$, we would recover the ordinary static lensing estimator~\cite{HuOkamoto:MassReconstruction2002}. There is an important difference in the mode coupling that arises in the case of the time-dependent deflection and that of static deflection.  In the limit where we utilize small-scale temperature modes to reconstruct large scale deflection potentials ($l_1 \gg L$) such that $\tilde{\mathcal{C}}_{l_1,0}^\mathcal{TT} \simeq \tilde{\mathcal{C}}_{l_2,0}^\mathcal{TT}$, the mode coupling in the static case is proportional to the large-scale convergence $f^\Gamma(\lvec_1,\Lvec-\lvec_1,0,0)\simeq L^2\tilde{\mathcal{C}}_{l_1,0}^\mathcal{TT}$ such that $\langle \mathcal{T}\mathcal{T} \rangle\propto L^2 \Gamma(L)$, whereas in the time-dependent case the mode coupling is proportional to the large-scale deflection $f^\Gamma(\lvec_1,\Lvec-\lvec_1,0,\Omega\neq0)\simeq (\Lvec\odot\lvec_1)\tilde{\mathcal{C}}_{l_1,0}^\mathcal{TT}$ such that $\langle \mathcal{T}\mathcal{T} \rangle\propto L \Gamma(L)$.  As a result, the reconstruction noise power is flat in convergence units for the static reconstruction, and flat in deflection units for the time-dependent reconstruction case. This distinction could be anticipated physically by noting that a static and spatially constant deflection  is unobservable, though a time-dependent deflection can be observed, even if it is spatially constant\footnote{We thank the anonymous referee for pointing out this important fact.}.   The different behavior of the reconstruction noise power on large scales is apparent below in Figure~\ref{fig:NLdd}.
 In the case of white noise in temperature, we find 
$\mathcal{N}_{L,\Omega}^{\phi\phi}=\mathcal{N}_{L,\Omega}^{\psi\psi}$.
The estimators for $\phi$ and for $\psi$ are orthogonal since the filters for the two estimators have opposite parity~\cite{Namikawa:2011cs}.

\begin{figure}
    \centering
    \includegraphics[width=\columnwidth]{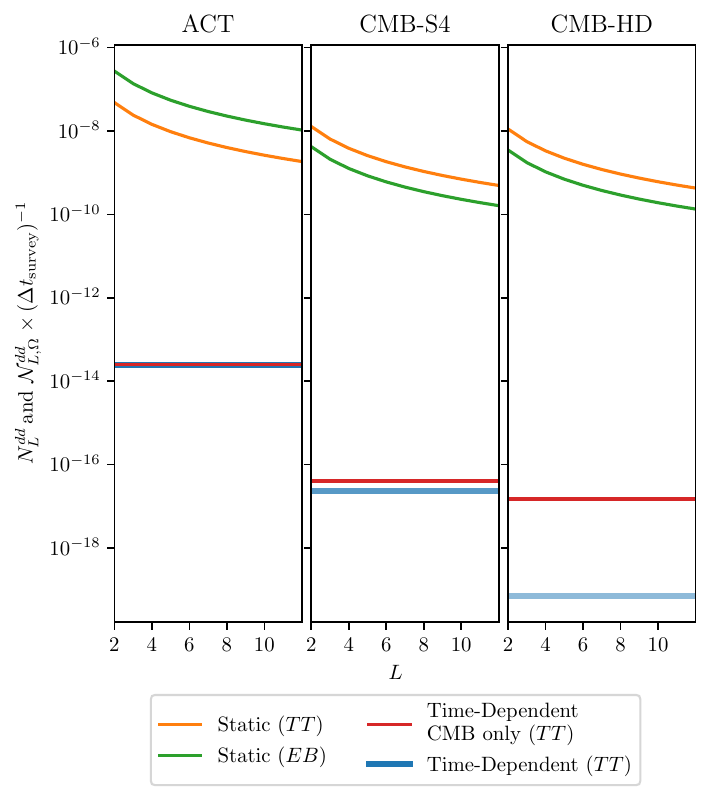}
    \caption{
    Forecasts for reconstructed deflection noise for CMB surveys with configurations like ACT (\textit{Left Panel}),  CMB-S4 (\textit{Middle Panel}), and  CMB-HD (\textit{Right Panel}). 
    The orange and green lines represent the deflection reconstruction noise using the standard $TT$ and $EB$ static quadratic lensing estimators~\cite{HuOkamoto:MassReconstruction2002}.
    For our time-dependent deflection reconstruction we plot $\tsurvey^{-1}\mathcal{N}_{L,\Omega}^{dd}$, which represents the deflection reconstruction noise in a frequency band of width of $\tsurvey^{-1}$, shown in red and blue.
    The red line shows the reconstruction noise  assuming no astrophysical foregrounds, while the blue line is the time-dependent deflection noise obtained from our estimator when including astrophysical foregrounds. 
    We focus on reconstruction of only the largest angular scales, since a stochastic background of gravitational waves produces primarily quadrupolar deflection, with power falling rapidly at smaller scales~\cite{BookFlanagan:AstrometricEffects2011}. 
    As can be seen, our time-dependent deflection  estimator exhibits significantly lower noise within each frequency band than the noise of standard static lensing quadratic estimator, and including astrophysical foregrounds improves rather than hampers the time-dependent deflection reconstruction.
    }
    \label{fig:NLdd}
\end{figure}

In contrast to standard CMB lensing estimators constructed from the weighted product of two modes both at $\Omega = 0$, our approach correlates the known CMB signal at $\Omega = 0$ with the induced fluctuation at $\Omega \neq 0$. 
As a result, the time-dependent deflection estimator benefits from the fact that the instrumental noise is uniform across all frequencies, while the primary CMB fluctuations reside only at $\Omega = 0$, meaning there is effectively no cosmic variance on one leg of the quadratic estimator.
These properties result in a significantly lower noise for the time-dependent deflection estimator than for the static deflection estimator.

One might have also expected that the astrophysical foregrounds which pose a significant challenge for various cosmological analyses would make the extraction of subtle time-dependent deflections challenging. 
However, astrophysical sources at distances greater than a GW wavelength experience the same deflection as primary CMB fluctuations, and therefore any astrophysical foregrounds that are static over the duration of the survey serve as useful sources that can be utilized in our time-dependent deflection reconstruction.  
In effect, we are agnostic to what contributes to the time-independent background that is deflected by the time-dependent gravitational waves, and the fact that astrophysical foregrounds can be resolved with high signal-to-noise ratios to smaller scales than can primary CMB fluctuations means that including foregrounds leads to lower rather than higher time-dependent deflection reconstruction noise.  Thus, in addition to the primary CMB, our estimator effectively uses the infrared, radio, and Sunyaev-Zel'dovich backgrounds as deflection sources (e.g., \citep{Schaan:2018yeh}).

In Figure~\ref{fig:NLdd}, we show the time-dependent deflection reconstruction 3d noise power in a frequency band of width $\tsurvey^{-1}$
 and the  static deflection reconstruction 2d noise power
$N_L^{dd}$ (which 
is the same as the time-dependent deflection 3d noise power in a frequency band of width $\tsurvey^{-1}$ 
at $\Omega=0$, $N_L^{dd}=\tsurvey^{-1}\mathcal{N}_{L,0}^{dd}$), and we have defined the deflection noise power in terms of the deflection potential noise power as $\mathcal{N}_{L,\Omega}^{dd}=L^2\mathcal{N}_{L,\Omega}^{\phi\phi}=L^2\mathcal{N}_{L,\Omega}^{\psi\psi}$.  We obtain these by evaluating Eq.~\eqref{eq:Nphi=NPsi}, using the \texttt{symlens} code~\cite{symlenscode} with our custom estimator.  (This expression is obtained in the flat sky formalism, but we extrapolate down to $L = 2$ using the fact that the time-dependent deflection noise is independent of $L$ on large scales.) Since the noise power is independent of $\Omega$---i.e., since the temporal white noise in the CMB leads to white noise in the reconstructed deflections---we implement the 2d version of the integrals, namely, the last line of Eq.~\eqref{eq:Nphi=NPsi}.

For the time-dependent deflection, we show cases both including and excluding astrophysical foregrounds as part of the static signal.  We assume that the undeflected, static CMB temperature is given by a power spectrum
\begin{equation}
\tilde{C}_l^{TT} = \tilde{C}_l^{TT,\mathrm{CMB}} + \tilde{C}_l^{TT,\mathrm{fg}}  \, .
\end{equation}
We take the foreground power spectrum to be a constant in $l$ for simplicity, since on small angular scales the emission is dominated by the Poisson nature of mm-wave galaxies.  Specifically, we assume a white noise-like power spectrum of $\tilde{C}_l^{TT,\mathrm{fg}}=(8~\mu$K-arcmin)$^{2}$, with this amplitude based on measurements of the cosmic infrared background at 150~GHz \cite{Dunkley:2010ge,ACT:2025fju}.

We show forecasts for the assumed CMB noise power $N_l^{TT}$ corresponding to 
ACT~\cite{ACT:2025fju,ACT:2025xdm} (treated with $\Delta_T=17.3~\mu$K-arcmin, $\theta_\mathrm{fwhm}=1$~arcmin, $\tsurvey=5$~yr\footnote{We use $\tsurvey=5$~yr since ACT spent 5 years of its 17-year lifespan on wide-field observations~\cite{ACT:2025xdm}.}), 
CMB-S4~\cite{CMB-S4:2016ple} (treated with $\Delta_T=0.71~\mu$K-arcmin, $\theta_\mathrm{fwhm}=1$~arcmin, $\tsurvey=10$~yr),
and CMB-HD~\cite{Sehgal:2019ewc} (treated with $\Delta_T=0.43~\mu$K-arcmin, $\theta_\mathrm{fwhm}=0.25$~arcmin, $\tsurvey=10$~yr). 
As demonstrated in Figure~\ref{fig:NLdd} the estimator of the time-dependent deflections performs substantially better within each frequency band of width 
$\tsurvey^{-1}$
than does static lensing reconstruction.
This can be understood from Eq.~\eqref{eq:Nphi=NPsi}, which shows that for $\Omega\neq0$, one leg of the estimator exhibits only the effects of instrumental noise with no cosmic variance.
Furthermore,
the proposed low-noise, high-resolution future surveys perform even better when leveraging the foregrounds, since they serve as a resolvable background for the time-dependent deflection due to GWs.

We could formulate our time-dependent deflection reconstruction using different polarization combinations as well. 
However, we focused here on the $TT$ estimator for time-dependent deflection since it has better performance than estimators constructed  from $E$ and $B$.
This is due to the fact that the time-dependent estimator benefits from having as many high signal-to-noise modes as possible on the $\omega=0$ leg, and as low a noise as possible on the $\Omega\neq0$ leg.
The time-independent signal for $T$ is significantly larger than $E$ or $B$ (from both primary CMB fluctuations and from astrophysical foregrounds), and the noise is also lower in $T$ for the small-scale CMB observations we consider here.
Furthermore, due to the angular dependence of the $EB$ mode coupling, the time-dependent deflection noise from the $EB$ reconstruction is flat in convergence units rather than deflection units on large angular scales, just as in the static case.
Therefore, unlike for static lensing reconstruction, where the $EB$ estimator is expected to provide the best performance at low noise (owing to a similar lack of cosmic variance in the unlensed $B$ field)~\cite{HuOkamoto:MassReconstruction2002}, the $TT$ estimator exhibits the best performance for time-dependent deflection reconstruction.

\section{Estimating the Amplitude of Gravitational Waves}
\label{sec:GW_Estimation}

We have seen that the time-dependent deflection reconstruction noise in any frequency band of width
$\tsurvey^{-1}$
is many orders of magnitude smaller than the ordinary lensing reconstruction noise.  
We now turn to quantifying the sensitivity of time-dependent reconstruction to a SGWB. 
The stochastic nature of the background, defined by $\Omega_\mathrm{GW}(f)$, predicts not a specific pattern of deflections $\mathbf{d}(\nhat,t)$, but rather the statistical deflection power as a function of angular scale and frequency 
$L^2\mathcal{C}_{L,\Omega}^{\Gamma\Gamma}$.
This is analogous to how a specific set of cosmological parameters determines the CMB power spectrum $C_l^{TT}$ and not the map of temperature fluctuations $T(\nhat)$.
We therefore need to determine how to measure $\Omega_\mathrm{GW}(f)$ from our estimate of the time-dependent deflection potentials $\hat{\Gamma}(\Lvec,\Omega)$, where $\Omega=2\pi f$.

As shown in Eq.~\eqref{eq:thetaQrmsl}, the deflection power for each value of $L$, $\Gamma$, and $f$ is given by
\begin{equation}
    \frac{L}{2\pi} L^2 2\pi f \mathcal{C}_{L,(2\pi f)}^{\Gamma\Gamma} =\frac{g_\Gamma \alpha_L^{\Gamma\Gamma}H_0^2 \Omega_\mathrm{GW}(f)}{4\pi^2f^2} \, ,
\end{equation} 
where the factor $L^2$ converts the lensing potential power to deflection power, the factor $\frac{L}{2\pi}$ relates the deflection power to the deflection variance, and the factor $2\pi f$ accounts for the fact that the power is defined per unit frequency, while $\Omega_\mathrm{GW}(f)$
gives the energy density per logarithmic frequency interval.
It is straightforward to check that summing over $L$ and $\Gamma$ and integrating over $\log f$ on both sides gives $\theta_\mathrm{rms}^2$.
Our reconstruction $\hat{\Gamma}(\Lvec,\Omega)$ allows us to form an estimate of the deflection potential power 
$\hat{\mathcal{C}}_{L,(2\pi f)}^{\Gamma\Gamma}$ 
(obtained for example by computing the power spectrum of the reconstructed map $\hat\Gamma$ and subtracting off the noise bias 
$\mathcal{N}_{L,(2\pi f)}^{\Gamma\Gamma}$), and this estimate has variance (e.g., \cite{Knox:1995dq})
\begin{equation}
    \mathrm{Var}\left[\hat{\mathcal{C}}^{\Gamma\Gamma}_{L,(2\pi f)}\right]=\frac{2 }{(2L+1)f_\mathrm{sky} } \left(\mathcal{N}_{L,(2\pi f)}^{\Gamma\Gamma}\right)^2 \, .
    \label{eq:GammaPowerVariance}
\end{equation}
Each value of $L$ and $\Gamma$ then provides an independent estimate of the GW energy density at each frequency
\begin{equation}
     \left[\hat\Omega_\mathrm{GW}(f)\right]_{L,\Gamma} =\frac{4\pi^2 L^3 f^3 }{g_\Gamma \alpha_L^{\Gamma\Gamma}H_0^2}\hat{\mathcal{C}}^{\Gamma\Gamma}_{L,(2\pi f)} \, ,
    \label{eq:OmegaGWinCGammaGamma}
\end{equation}
which can be combined into a single unbiased, minimum-variance estimator $\left[\hat\Omega_{\mathrm{GW}}(f)\right]_\mathrm{mv}$
using an inverse-variance weighting of the independent estimators.
The variance of the combined estimator is then
\begin{align}
    &\mathrm{Var} \left[ \left[\hat\Omega_\mathrm{GW}(f)\right]_\mathrm{mv} \right]  \nonumber \\
    & \quad = \left[\sum_{\Gamma=\phi,\psi}\sum_{L=2}^{\infty} \frac{(2 L + 1)f_\mathrm{sky}}{2} \left(\frac{g_{\Gamma} \,\alpha^{\Gamma \Gamma}_L H_0^2}{4\pi^2 f^3 L^3 \mathcal{N}^{\Gamma\Gamma}_{L,(2\pi f)}}\right)^2\right]^{-1} \,\nonumber \\
    & \quad \simeq {147456 \pi^4 \over 125} {f^6 \over H_0^4} {1\over f_\mathrm{sky}} \left( \mathcal{N}_{2,(2\pi f)}^{\phi\phi} \right )^2
    \, ,
\end{align}
where in the last line we approximated that the signal is dominant at $L=2$, such that $\alpha_2^{\Gamma\Gamma} = 5/6$~\cite{BookFlanagan:AstrometricEffects2011,PyneEtAl:GravitationalRadiation1996} and the other $\alpha_L^{\Gamma\Gamma}$ are much smaller; and we used the fact that curl and gradient modes have equal contribution, so that $g_\phi = g_\psi = {1\over 2}$.

\begin{figure}
    \centering
    \includegraphics[width=\columnwidth]{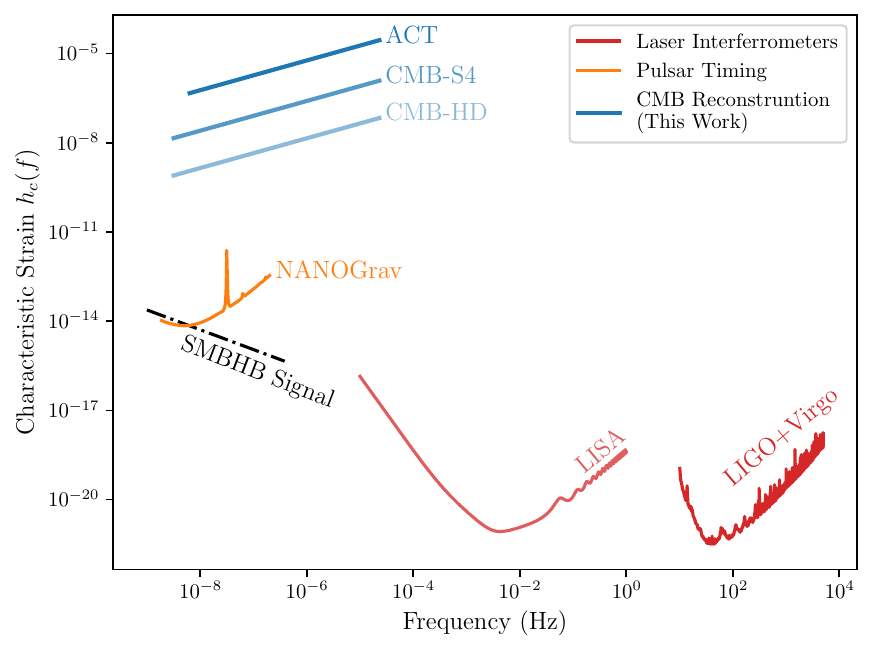}
    \caption{
    Sensitivity of various methods to search for gravitational waves across a wide range of frequencies.
    The forecasted sensitivity of our new time-dependent deflection reconstruction method applied to current and future CMB surveys is shown in blue.
    The existing sensitivity from pulsar timing array analysis by NANOGrav~\cite{NANOGrav:2023hfp} is shown in orange.
    Laser interferometer sensitivity from existing ground-based observations (LIGO+Virgo)~\cite{KAGRA:2023pio} and future space-based observations (LISA)~\cite{LISA:2024hlh} are shown in red.
    The spectrum of a stochastic background of gravitational waves produced by a population of supermassive black hole binaries, with an amplitude matching the inferred value from NANOGrav observations~\cite{NANOGrav:2023hfp}, is shown as a black dash-dot line.
    Although the CMB reconstruction is not competitive with pulsar timing in their common frequency band, it does provide sensitivity in the microhertz frequency gap between pulsar timing (NANOGrav) and future space-based laser interferometer (LISA) observations.
    }
    \label{fig:Senscurve}
\end{figure}

It is conventional to express the GW power in terms of the characteristic strain $h_c(f)$ or the power spectral density $S_h(f)$, which are related to the GW energy density as 
\begin{equation}
     H_0^2 \Omega_{\mathrm{GW}}(f)
     =\frac{2\pi^2}{3}f^2 [h_c(f)]^2=\frac{2\pi^2}{3}f^3 S_h(f) \, ,
    \label{eq:conversions}
\end{equation}
with similarly defined quantities $h_n(f)$ and $S_n(f)$ describing the noise characteristic strain and noise power spectral density~\cite{MooreEtAl:GravitationalwaveSensitivity2015}.
In Figure~\ref{fig:Senscurve}, we show forecasts for the sensitivity to a SGWB, in terms of the characteristic strain $h_c(f)$, using our new time-dependent deflection reconstruction technique for the CMB surveys ACT, CMB-S4, and CMB-HD, together with those from standard techniques. 
While the sensitivity in the nanohertz regime of PTAs significantly exceeds what can be achieved with our time-dependent reconstruction estimator applied to CMB surveys, our technique extends into a frequency range above that typically measured by PTAs and below the range that will be probed by LISA, in the microhertz gap.

The range of GW frequencies that is resolvable by our time-dependent reconstruction method is set on the low end by the duration of the survey, $\tsurvey$, and on the high end by the cadence with which we can make maps of the sky, $\tcadence$, giving $f \in \left[\left(2\times10^{-9}~\mathrm{Hz}\right)\frac{10 \,\mathrm{yr}}{\tsurvey}, \left(2\times10^{-5}~\mathrm{Hz}\right)\frac{1 \, \mathrm{day}}{\Delta t_\mathrm{cadence}}\right]$. 
Surveys of the CMB routinely cover large fractions of the sky each day, thereby allowing for $\tcadence\sim1$~day without any change to survey strategy. 
Furthermore, past and planned CMB surveys have long durations, $\tsurvey\sim$10~yr, thereby naturally allowing for a broad frequency coverage with our time-dependent deflection reconstruction.

One of the natural sources of GWs  within the relevant frequency range is the SGWB originating from a population of supermassive black hole binaries (SMBHBs)~\cite{Sesana:2008mz}, which produce GWs in the nanohertz to microhertz regime.
The NANOGrav collaboration has recently found evidence for a SGWB of the form
\begin{equation}
    h_c(f)=A_{\mathrm{GWB}} \left(\frac{f}{f_0}\right)^{-\frac{\gamma-3}{2}} \, ,
    \label{eq:signal}
\end{equation}
where $\gamma=13/3$ for SMBHBs, with a reference frequency of $f_{0} = 1~\mathrm{yr}^{-1} = 3\times10^{-8}~\mathrm{Hz}$, and an amplitude $A_{\mathrm{GWB}} = 2.4_{-0.6}^{+0.7}\times 10^{-15}$~\cite{NANOGrav:2023hfp}.  We plot this spectrum along with the sensitivity curves in Figure~\ref{fig:Senscurve}.

\section{Forecasts}
\label{sec:Forecasts}
\subsection{Direct Constraint}

Next, we can calculate the signal-to-noise ratio with which our time-dependent deflection reconstruction can observe the SGWB.  The square of the signal-to-noise for a persistent signal is given by~\cite{HazbounEtAl:RealisticSensitivity2019,MooreEtAl:GravitationalwaveSensitivity2015,ThraneRomano:SensitivityCurves2013}
\begin{equation}
    \mathrm{SNR}^2=2\tsurvey \int \left(\frac{S_h(f)}{S_n(f)}\right)^2 df \, ,
\end{equation}
thereby giving
\begin{equation}
    \mathrm{SNR}^2=2\tsurvey \int \frac{\Omega_\mathrm{GW}(f)^2}{\mathrm{Var}\left[\left[\hat\Omega_\mathrm{GW}(f)\right]_\mathrm{mv}\right]} df \, .
\end{equation}
For a SGWB resulting from SMBHBs consistent with the NANOGrav signal, we estimate $\mathrm{SNR}=1.7\times10^{-16}$ for ACT, $\mathrm{SNR}=4.5\times10^{-13}$ for CMB-S4, and $\mathrm{SNR}=1.4\times10^{-10}$ for CMB-HD (where each of these estimates includes the improvement that comes from considering the deflection of astrophysical foregrounds as well as the primary CMB, and assumes $f_\mathrm{sky}=0.5$).  This unfortunately puts a detection of the SGWB from SMBHBs via time-dependent deflection reconstruction applied to the CMB out of reach for current and proposed CMB surveys.

We can understand how the SNR scales with the experimental configuration as follows.  First, we can express our time-dependent deflection reconstruction noise at low $L$ where the signal from GWs is non-negligible as
\begin{align}
    \mathcal{N}^{\Gamma\Gamma}_{L,\Omega} 
    &= \tsurvey\left[ \int \dtwol{1} \left(\lvec_1 \odot \Lvec \right)^2 \, \frac{\left(\tilde C^{TT}_{l_1}\right)^2}{ C_{l_1}^{TT,\mathrm{obs}} N_{\left|\Lvec-\lvec_1\right|}^{TT}}  \right]^{-1} \nonumber \\
    &\sim \tsurvey \left[ \int_0^{l_\mathrm{max}} dl \,  l^3 \left(\frac{\tilde{C}_l^{TT}}{N_l^{TT}}\right)  \right]^{-1} \nonumber \\
    & \sim \tsurvey \theta_\mathrm{fwhm}^4 \left(\frac{\Delta_T^2}{\tilde{C}_l^{TT,\mathrm{fg}}}\right)  \\
    & \sim  \theta_\mathrm{fwhm}^4 \left(\frac{\mathrm{NET}_\mathrm{survey}^2 f_\mathrm{sky}}{\tilde{C}_l^{TT,\mathrm{fg}}}\right)\, .
\end{align}
In this expression, we have used the fact that the estimator utilizes a Wiener filter of the temperature on the $\omega=0$ leg, thereby effectively cutting off the integral at $l_\mathrm{max}$, defined as the largest $l$ where $(\tilde{C}_l^{TT,\mathrm{CMB}}+\tilde{C}_l^{TT,\mathrm{fg}})/C_l^{TT,\mathrm{obs}}\sim \mathcal{O}(1)$. For cases when the white noise power is below the foreground power $N_l^{TT}<\tilde{C}_l^{TT,\mathrm{fg}}$, we can approximate $l_\mathrm{max}$ as the beam scale $l_\mathrm{beam}=\frac{\sqrt{8\log 2}}{\theta_\mathrm{fwhm}}$. The time-dependent leg of the estimator measures the signal of the observed time-independent background divided by the instrumental noise, which can be estimated as the signal-to-noise of the foregrounds, since they provide the main contribution to the power on the small scales that dominate the integral 
($\tilde{C}_l^{TT,\mathrm{CMB}}\ll \tilde{C}_l^{TT,\mathrm{fg}}$ for $l \sim l_\mathrm{max}$), allowing us to approximate
$(\tilde{C}_l^{TT}+\tilde{C}_l^{\mathrm{fg}})/N_l^{TT}\sim \frac{\tilde{C}_l^\mathrm{fg}}{\Delta_T^2}$.
Next, since the deflection reconstruction power is independent of frequency (for $\Omega\neq0$), we can perform the frequency integral analytically to find the dependence on the survey duration
\begin{align}
    \mathrm{SNR}^2&\propto\tsurvey \int \left(\frac{S_h(f)}{S_n(f)}\right)^2 df \nonumber \\
    &\propto \tsurvey \int_{\tsurvey^{-1}}^{\tcadence^{-1}} \frac{f^{-2\gamma+4}}{\tsurvey^2 f^6} df \nonumber \\
    &\propto (\tsurvey)^{2\gamma}  \, ,
\end{align}
where we have assumed that $\tsurvey \gg \tcadence$.
Putting this together we find
\begin{equation}
    \mathrm{SNR} \propto (\tsurvey)^{\gamma} \theta_{\mathrm{fwhm}}^{-4} \Delta_T^{-2} \tilde{C}_l^{TT,\mathrm{fg}} \, ,
\end{equation}
or, restoring order-one factors,
\begin{align}
    \mathrm{SNR}& \approx
    4\times10^{-13} \left(\frac{1\, \mathrm{arcmin}}{\theta_\mathrm{fwhm}}\right)^4\\ \nonumber 
    & \hphantom{{}=} \times \left(\frac{0.7~ \muK\mhyphen\mathrm{arcmin}}{\Delta_T}\right)^2 \left(\frac{\tilde{C}_l^{TT,\mathrm{fg}}}{(8\, \muK\mhyphen\mathrm{arcmin})^2}\right)\\ \nonumber 
    &\hphantom{{}=} \times \left(\frac{\tsurvey}{10\,\mathrm{yr}}\right)^{\gamma}\left(\frac{A_\mathrm{SGWB}}{2.4\times10^{-15}}\right)^2 \, .
\end{align}
This approximates the SNR within about a factor of 2 for the configurations we used above.

Alternatively, the most natural way to frame the noise properties may be in terms of the instantaneous sensitivity of the survey, since temporal Fourier modes of the noise correspond directly to the noise of the time-dependent deflection reconstruction at that same frequency.  Assuming a sky coverage of  $f_\mathrm{sky}=0.5$, this yields
\begin{align}
    \mathrm{SNR}&\approx
    4\times10^{-13} \left(\frac{1\, \mathrm{arcmin}}{\theta_\mathrm{fwhm}}\right)^4\\ \nonumber 
    &\hphantom{{}=} \times \left(\frac{1.5~ \muK\sqrt{s}}{\mathrm{NET}_\mathrm{survey}}\right)^2 \left(\frac{\tilde{C}_l^{TT,\mathrm{fg}}}{(8\, \muK\mhyphen\mathrm{arcmin})^2}\right)\\ \nonumber 
    &\hphantom{{}=} \times \left(\frac{\tsurvey}{10\,\mathrm{yr}}\right)^{\gamma+1}\left(\frac{A_\mathrm{SGWB}}{2.4\times10^{-15}}\right)^2 \, ,
\end{align}
where we have used $\mathrm{NET}_\mathrm{survey}=\frac{\Delta_T \sqrt{\tsurvey}}{\sqrt{4\pi f_\mathrm{sky}}}$.
This sensitivity is unfortunately far below what is currently achieved with PTAs~\cite{NANOGrav:2023hfp}, but it is comparable to recent estimates of the sensitivity from cosmic shear surveys~\cite{MentastiContaldi:CosmicShimmering2024}.

\subsection{Integrated constraint}

While the sensitivity of time-dependent deflection reconstruction applied to the CMB is not competitive with PTAs in the nanohertz band, it does offer some advantages.
Typical PTA searches for GWs are not sensitive to very low frequencies, because the timing delays generated by GWs with periods exceeding the survey duration are degenerate with parameters used in the pulsar timing model~\cite{Moore:2014eua} (although some efforts have been made to use the induced parameter drifts to probe the low-frequency regime with PTAs as well~\cite{DeRocco:2022irl}).
Like other astrometric searches for gravitational waves, the method we describe here provides constraints to very low frequency GWs.  In principle, the sensitivity to these ultra-low frequency GWs allows the search for the SGWB predicted by dark sector phase transitions~\cite{MooreVecchio:UltralowFrequency2021}, turbulent QCD phase transitions~\cite{NeronovEtAl:NANOGravSignal2021,BrandenburgEtAl:CanWe2021}, or by cosmic strings~\cite{Vilenkin:1981bx,Vachaspati:1984gt,Caprini:2018mtu,Buchmuller:2021mbb}.

The deflections caused by GWs with periods longer than the duration of the survey will cause a secular drift in astrophysical source positions.  
The contribution to the mean square angular velocity due to a GW of frequency $f$ is related to the mean square deflection by
$\dot{\theta}_\mathrm{rms}^2(f) = 4\pi^2f^2\theta_\mathrm{rms}^2(f)$,
and the observed angular velocity receives contributions from all frequencies~\cite{PyneEtAl:GravitationalRadiation1996,BookFlanagan:AstrometricEffects2011}.
The mean square angular velocity of source positions due to GWs below the resolvable band is
\begin{equation}
    \left(\dot{\theta}_\mathrm{rms}^\mathrm{low}\right)^2 =  \int_{f<\tsurvey^{-1}} d\log f \, H_0^2 \Omega_\mathrm{GW}(f) \, , 
\end{equation}
such that the angular velocity power spectrum from these low frequency GWs is
\begin{equation}
    4\pi^2L^3f^3\mathcal{C}_{L,(2\pi f)}^{\Gamma\Gamma} = g_\Gamma \alpha_L^{\Gamma\Gamma} \left(\dot{\theta}_\mathrm{rms}^\mathrm{low}\right)^2 \, .
    \label{eq:LowFrequencyDeflection}
\end{equation}
This
indicates how the contribution to the angular velocity from GWs below the resolvable range of frequencies are aliased into the resolvable band in each log frequency bin.

We can then construct an independent estimate of the integrated GW density at low frequency $\Omega_\mathrm{GW}^\mathrm{low} \equiv \int_{f<\tsurvey^{-1}} d\log f \,{ \Omega_\mathrm{GW}(f)}$ for each value of $L$, $\Gamma$, and $f$ as
\begin{equation}
    \left[\hat{\Omega}_\mathrm{GW}^\mathrm{low}\right]_{L,\Gamma,f} = \frac{4\pi^2 L^3 f^3 }{g_\Gamma \alpha_L^{\Gamma\Gamma}H_0^2}\hat{\mathcal{C}}^{\Gamma\Gamma}_{L,(2\pi f)} \, ,
    \label{eq:Integrated_Constraint}
\end{equation}
which we can see is the same as the estimator for $\Omega_\mathrm{GW}(f)$ for a resolvable GW frequency shown in Eq.~\eqref{eq:OmegaGWinCGammaGamma}.  We can therefore interpret upper limits on $\Omega_\mathrm{GW}(f)$ also as upper limits on the integrated low-frequency GW spectrum $\Omega_\mathrm{GW}^\mathrm{low}$.
The variance of the estimator of the low-frequency GW density after summing over all observable logarithmic frequency bins via inverse-variance weighting is then
\begin{align}
    &\mathrm{Var} \left[ \left[\hat{\Omega}_\mathrm{GW}^\mathrm{low}\right]_\mathrm{mv} \right] =
    \left[\int_{f=\tsurvey^{-1}}^{f=\tcadence^{-1}} d\log f  %
    \vphantom{\left(\frac{g_{\Gamma} \,\alpha^{\Gamma \Gamma}_L H_0^2}{4\pi^2 f^3 L^3 \mathcal{N}^{\Gamma\Gamma}_{L,(2\pi f)}}\right)^2}\right.
    \nonumber \\
    & \quad \times \sum_{\Gamma=\phi,\psi}\sum_{L=2}^{\infty} \frac{(2 L + 1)f_\mathrm{sky}}{2}
    \left. \left(\frac{g_{\Gamma} \,\alpha^{\Gamma \Gamma}_L H_0^2}{4\pi^2 f^3 L^3 \mathcal{N}^{\Gamma\Gamma}_{L,(2\pi f)}}\right)^2\right]^{-1}  .
\end{align}

Using the same configurations as above, we forecast that ACT, CMB-S4, and CMB-HD could place constraints on the integrated low-frequency GW spectrum of $\Omega_\mathrm{GW}^\mathrm{low}<2\times10^8$, $\Omega_\mathrm{GW}^\mathrm{low}<6\times10^4$, and $\Omega_\mathrm{GW}^\mathrm{low}<2\times10^2$, respectively.
These constraints are unfortunately not particularly useful, since $\Omega_\mathrm{GW}^\mathrm{low}=1$ would imply that the total energy density of the present universe is dominated by very low frequency GWs, which is obviously in conflict with observed cosmological evolution.
Current constraints from astrometric measurements of quasars with \textsl{Gaia} set an upper limit of $\Omega_\mathrm{GW}^\mathrm{low}<0.0096$~\cite{Darling:2024myz}, and the full \textsl{Gaia} data set may reduce this constraint to $\Omega_\mathrm{GW}^\mathrm{low}<0.0006$~\cite{Darling:2018hmc}.
Another relevant point of comparison is the constraint on the SGWB energy density that comes from cosmological measurements of the light relic density $N_\mathrm{eff}$~\cite{Boyle:2007zx,Meerburg:2015zua,Lasky:2015lej}.  This quantity can be measured with the abundance of light elements produced during big bang nucleosynthesis or through the impact of radiation on the CMB~\cite{Dvorkin:2022jyg}. However, $N_\mathrm{eff}$ only receives contributions from GWs present around the time of big bang nucleosynthesis or recombination and is sensitive to a different range of frequencies than can be probed via astrometry~\cite{MooreVecchio:UltralowFrequency2021}.

\section{Discussion and Conclusion}
\label{sec:Conclusion}

Gravitational waves passing through Earth impart time-dependent deflections on observed astrophysical source positions.
This physics provides one example of a time-dependent signal that we might hope to isolate with current and future cosmological surveys.
CMB surveys map large fractions of the sky at a high cadence, offering an opportunity to search for such time-dependent signals.
In this paper, we constructed a new time-dependent deflection estimator that allows us to search for GWs using CMB observations.

Searching for time-dependent signals in survey data can be achieved in a number of different ways.
One simple method is to compare maps constructed from observations taken at distinct times and analyze their differences.
Another method, that we advocate here, relies on applying a Fourier transform over the time domain.
This allows for the time-independent components to be separated and provides a set of maps at non-zero temporal frequency with the same noise properties as the time-averaged, or zero-frequency map, that is normally used in science analyses.
Furthermore, the cosmic variance from the time-independent components is restricted to the zero-frequency map.
Through this construction, we developed a quadratic estimator to search for time-dependent deflections of the CMB.

The estimator we constructed works by correlating the time-independent CMB map with that at nonzero frequency, in order to reconstruct defections that fluctuate over that nonzero frequency.
We found that it exhibits much lower noise than its time-independent counterpart (the familiar lensing quadratic estimator~\cite{HuOkamoto:MassReconstruction2002}) for the same experimental configuration.
We attributed the lower noise to the lack of cosmic variance on the non-zero frequency leg of the estimator.
This estimator can be applied to existing and future surveys without any modification to experimental design or survey strategy, since CMB surveys are already being constructed to search for short-timescale transient events, over large sky fractions, and with high angular resolution.

We showed how our time-dependent deflection estimator could be applied to search for GWs using current and planned CMB observations.
Since GWs passing through the Earth deflect all astronomical source positions, we demonstrated that including astrophysical foregrounds could be included as part of the deflected source in our analysis, leading to an even smaller deflection noise.
We estimated the sensitivity of various CMB surveys to the SGWB (which has a frequency range within temporal parameters of CMB surveys) using our time-dependent deflection reconstruction method.
Unfortunately, the sensitivity to GWs provided by proposed CMB surveys is unlikely to be competitive with PTA searches for GWs within their common frequency band.

The time-dependent deflection reconstruction method we developed here shares many of the advantages of astrometric searches for GWs, which are highly complementary to PTA searches.
Since angular deflections of astrophysical sources observed in astrometric searches and the timing residuals observed by PTAs are generated by the same GWs passing through Earth, one can make definite predictions about the correlations among these observables~\cite{Qin:2018yhy}.
Combined analysis of angular deflections and timing residuals measured allows further insights into the SGWB~\cite{Cruz:2024diu}, including searches for chirality of the GW background and tests of general relativity~\cite{Qin:2018yhy,Qin:2020hfy,Caliskan:2023cqm,MentastiContaldi:CosmicShimmering2024}.

The high cadence with which CMB surveys can map the sky allows time-dependent deflection reconstruction to be used to search for GWs in the microhertz band, a frequency range that has been challenging to probe by other means.
The current best constraints on GWs in microhertz band come from Doppler tracking of the \textit{Cassini} satellite~\cite{Armstrong:2003ay}, while the millihertz band is currently best constrained by monitoring the Earth's normal vibrational modes~\cite{Coughlin:2014xua}.
Dramatic improvements in the millihertz band are expected from the LISA mission~\cite{LISA:2024hlh}, but the microhertz band remains a challenge.
Proposals have been made for GW searches in the microhertz regime using dedicated artificial spacecraft~\cite{Sesana:2019vho}, by outfitting asteroids with appropriate instrumentation~\cite{Fedderke:2021kuy}, through binary resonance searches utilizing laser ranging of the moon or artificial satellites or by monitoring millisecond pulsar binaries~\cite{Blas:2021mqw}, with high-cadence relative astrometric measurements using galaxy surveys~\cite{Wang:2020pmf,Wang:2022sxn,Pardo:2023cag}, and searching for modulation of continuous GWs detected by experiments like LISA~\cite{Bustamante-Rosell:2021daj}.

Similarly, the sub-nanohertz band is challenging to probe with PTA searches for GWs, since the timing residuals caused by GWs with periods longer than the survey duration (of roughly 20 years) are degenerate with some of the the parameters that define the pulsar timing model.
By contrast, astrometric searches for GWs as well as the time-dependent deflection reconstruction developed here, are sensitive to the integral of the low-frequency GW spectrum.
We showed how CMB surveys can be used to place a constraint on the integrated sub-nanohertz GW spectrum, though we found those constraints to be weaker than existing astrometric constraints derived from \textsl{Gaia} data.

The framework of time-dependent deflection reconstruction that we developed here can be straightforwardly extended.
Other time-dependent distortions of CMB surveys, such as instrumental gain calibrations, temperature-to-polarization leakages, or instrumental beams which vary in time, could be ``self calibrated" with the CMB data itself, using the framework we developed here---the static version of which (e.g.,~\cite{Su:2009tp,Yadav:2009za}) has been performed on real data~\citep{BICEPKeck:2022kci}.
New physics can also imprint time-dependent signals onto the CMB that can be isolated with the techniques described here.
For example, an oscillating axion field can lead to time-dependent polarization rotation that can be sought in the CMB~\cite{Fedderke:2019ajk,SPT-3G:2022ods,POLARBEAR:2023ric}
Furthermore, one can apply the techniques developed here to observations at other wavelengths, allowing for similar estimators constructed with optical galaxy surveys (as in Refs.~\cite{Wang:2020pmf,Wang:2022sxn,Pardo:2023cag,MentastiContaldi:CosmicShimmering2024}) or with line intensity mapping surveys (like a time-dependent version of the lensing estimators described in Refs.~\cite{Zahn:2005ap,Foreman:2018gnv}).

The time-dependent deflection reconstruction we describe here provides a new way to utilize the data that is being collected by CMB surveys.
Typical science analyses of CMB data use only the static signals present in the time-averaged maps constructed from the survey.
The techniques we developed here make use of the parts of the data that lie at non-zero temporal frequency, and which are often discarded in standard analyses.
Study of the transient and dynamic mm-wave sky has recently emerged as a target of CMB  surveys~\cite{Abazajian:2019eic,SimonsObservatory:2025wwn}, though the focus has so far primarily been on point-like sources.
We demonstrated here how a more complete treatment of time-dependent effects in CMB surveys may reveal phenomena that are not captured by other means.
We focused our attention on one such phenomenon---the deflection of CMB fluctuations by GWs passing through the Earth.
However, the general framework that we developed will be useful beyond this particular science goal, creating new opportunities to more fully capture the science available in the dynamic millimeter-wave sky.

\section*{Acknowledgments}

The authors thank Colin Hill, Mathew Madhavacheril, and Toshiya Namikawa for helpful discussions.
We also thank the anonymous referee who suggested a physical justification for the shape of the reconstruction noise curves in Figure~\ref{fig:NLdd}.
Computational resources for this research were provided by SMU’s O’Donnell Data Science and Research Computing Institute.
JM and AvE are supported by NASA through ADAP grant \mbox{80NSSC24K0665} and by NSF through grant \mbox{AST-2510926}.  JM is additionally supported by the US~Department of Energy under Grant~\mbox{DE-SC0010129}. AvE is additionally supported by NASA ADAP grants 80NSSC23K0747 and 80NSSC23K0464, and NSF AAG grant 588167.

\bibliographystyle{utphys}
\bibliography{gw_search}

\end{document}